# Effect of Na doping on flux pinning of YBa$_{1.9}$Na$_{0.1}$Cu$_3$O$_{7-\delta}$

S. Y. Ding*, X. Leng, L. Qiu, Z.H. Wang, H. Luo, Y. Liu
National Laboratory of Solid State Microstructures, Department of Physics,
Nanjing University, Nanjing 210093, P. R. China
H. Zhang,
Shanghai Institute of Metallurgy, Chinese Academy of Sciences,
Shanghai 200050, P. R. China

**Abstract**

We have prepared Na-doped YBa$_2$Cu$_3$O$_y$ (YBa$_{1.9}$Na$_{0.1}$Cu$_3$O$_y$ +40mol%Y211) (YBNCO) and Na-free YBa$_2$Cu$_3$O$_y$ (YBCO) samples by the Melt-Textured Growth (MTG) method to study the effect of doped Na ion on flux pinning. The ac susceptibility curves (acs) as well as the hysteresis loops were measured for the samples. Then the effective pinning energy (U(T,H$_{dc}$,J)), irreversibility line (H$_{irr}$(T)) and critical current density (j$_c$(H$_{dc}$)) were determined, where T, H$_{dc}$ and J are temperature, dc magnetic field and current density, respectively. We found that, with Na doping, the H$_{irr}$(T) line shifted to lower temperature while the J$_c$(H$_{dc}$) and U(T,H$_{dc}$,J) became smaller. It indicates that the Na ions play a negative role in the flux pinning of YBCO. The appearance of the second peak in the J$_c$(H$_{dc}$) curves and the enhancement of anisotropy in YBNCO further support this finding.

*To whom all correspondence should be addressed.
E-mail: syding@nju.edu.cn; Fax: +86-25-3595535;

# 1. Introduction

There have been number of papers reported on the Na substitution in $YBa_2Cu_3O_y$ compound [1-7]. However, most of the works concern the effect of Na ions on the oriented grain growth and critical temperature $T_c$ in the polycrystalline YBCO. These studies have shown that $Na^+$ favors both the $Y^{3+}$ as well as the $Ba^{2+}$ sites [1-3] since the ionic radius of $Na^+$ (0.095nm) is comparable to that of $Y^{3+}$ (0.093nm) and $Ba^{2+}$ (0.135nm). No change in $T_c$ has been observed when the doped concentration x < 0.2 in the $YBa_{2-x}Na_xCu_3O_y$ (YBNCO) [4-8].

MTG method provides a favorable condition for investigating the effects of chemical doping on the flux pinning of the MTG-YBCO (hereafter YBCO). Recent measurements, such as resistance and magnetization curves, have found that with certain Na doping, $J_c$ of YBCO was changed and the magnetization curves showed a second peak [6-8]. It was expected that flux pinning in YBCO might be improved by Na doping. However, both enhancement of flux pinning and appearance of the second peak in a same sample may result in confusion because the second peak usually takes place in the samples with weak flux pinning systems.

The aim of this work is to probe the true effect of Na addition on flux pinning in YBNCO mainly by acs experiment. Ac susceptibility is a powerful tool on flux pinning studies not only for its convenience but also for it facilitates to change the frequency f and the amplitude of the ac field. We have prepared Na-doped and Na-free samples by MTG process and measured the temperature dependence of acs at different applied dc and ac fields in addition to the measurements of magnetization. The field, temperature and current dependencies of effective activation energy, the irreversibility line and critical current density were determined.

## 2. Experiments and data analyze

The bulks YBNCO with apparent composition of $YBa_{1.9}Na_{0.1}Cu_3O_y+40mol\%Y_2BaCuO_5$ were prepared by the MTG method. The details of the method were reported elsewhere [7]. The crystal samples with platelet splitted up from the cleaved plane of the bulk with a single domain. The size is 4mm in diameter with 0.35mm thick. And the c axis is perpendicular to the wider surfaces. The acs was measured in an ac susceptometer with high sensitivity at different f, $h_{ac}$, and $H_{dc}$. The applied field is H (t)=$H_{dc}+h_{ac}\cos(2\pi ft)$, $H_{dc}//c$, $h_{ac}//c$, $H_{dc}>>h_{ac}$ or $H_{dc}$=0. The dc magnetization measurement was carried out by a VSM (vibrating sample magnetometer).

It is supposed that there is giant flux creep in the samples. It has been shown that for a thin slab with thickness 2d in applied field H (t) the effective pinning energy is [9]

$$U(J) = T\ln(1+t/t_0) = T\ln(1+1/ft_0) \approx -\ln(ft_o) \qquad (1)$$

where $t_o$ is the time scale and $t=1/f$. Thus the U (J) relationship can be determined experimentally by acs measurement at different frequency and current densities. Because of the highly non-linearity of flux creep, special constant but time dependent current density approximation (i.e. the expended Bean model) provides a good description of the field profile inside a sample [9]. The current density can be estimated directly as

$$J = h_{ac}/d, \qquad T = T_p \qquad (2)$$

where $T_p$ is the temperature at which the imaginary part of complex acs($\chi = \chi'-\chi''$), $\chi''$, peaks. Furthermore, assuming $U(J,H_{dc},T) = U_o(J,H_{dc})G(T)$ and combining (1) and (2) where $G(T_p)$ accounts for the temperature dependence of U, one can re-write equation(1) as [10]:

$$G(T_p)/T_p = [U_o(J,H_{dc})]^{-1} 1/(ft_0)) \qquad (3)$$

Equation (3) and (2) show that a systematic measurement of $\chi''$ peak as a function of $h_{ac}$, f and $H_{dc}$ can extract the effective pinning barriers $U(T, H_{dc}, J)$ [9,10].

## 3. Results and discussions
### 3.1. Irreversibility line

Fig.1 shows the effect of the dc magnetic field on the $\chi(T)$ curve of the YBNCO at the fixed frequency and amplitude of ac field. It is seen that as $H_{dc}$ increased from around 0.5T to 6T, the $\chi(T)$ curve shifts towards lower temperatures and the transition is broadened. On the other hand, for YBCO sample, the shift is smaller and the transition is broadening that are not illustrated here for simplicity. We define the transition temperature as the irreversibility temperature at $H_{dc}$, where acs onsets. In this way the so-called irreversibility line $T_c(H_{dc})$ or $H_{irr}(T)$ is obtained. Fig2 shows the irreversibility lines of YBCO and YBNCO for $H_{dc}$//c obtained by dc and ac magnetic measurements, respectively. From Fig2, we can find that the line of $H_{irr}(T)$ for YBNCO crystal is shifted to lower temperature, indicating the Na addition has depressed the flux pinning .

Shown in the inset in Fig.2 is also the influence of the driving frequency of ac field on the irreversibility line of YBNCO. It is seen that $H_{irr}(T)$ is frequency dependent which is in agreement with references [11]. The effect of amplitude of ac field on the irreversibility line was measured as well, and shown that $h_{ac}$ (i.e. the current density J) affect the irreversibility line too. These results indicate that the onset temperature $T_c$ ($H_{dc}$) is in fact determined by the flux dynamics, implying that our definition on the irreversibility temperature is reasonable. Hence not only the peak temperature $T_p$ of $\chi''$ [11] but also the onset temperature of acs can be used to define as the irreversibility lines.

### 3.2. Effective activated energy

Shown in Fig.3 are typical $\chi'(T)$ and $\chi''(T)$ curves for the YBNCO sample at $H_{dc}$=0.5T, $h_{ac}$=0.4Gs and different frequency, where we can see that $T_p$ increases with increasing f. To extract the effective barriers, the data of acs are fitted by

$$G(T_p) = [1 - (T_p/T_c)^1]^{1.5} \qquad (4)$$

where $T_c$ = 90 K, the onset temperature in zero field. That is to say, plotting the acs data as $G(T_p)/T_p$ versus lnf either at fixed $h_{ac}$ in fixed dc field $H_{dc}$ (see Fig.4) or at fixed $H_{dc}$ in different amplitude of ac field $h_{ac}$ (not shown here for simplicity), we should obtain a set of straight lines whose slop and intercept are the inverses of effective barriers $(U_o(J,H_{dc}))^{-1}$ and time scale $1/t_o$, respectively, see Fig.5 and, Fig.6. It is seen that the equations can fit the data well.

Displayed in Fig4 is a summary of the data (symbols) in different dc fields and frequency at fixed $h_{ac}$, showing the effect of dc fields $H_{dc}$ on the slopes $(U_o(J,H_{dc}))^{-1}$ and intercepts $1/t_0$ of the fitting straight lines. It is found that while $1/t_0$ only varies slightly with $H_{dc}$ around its mean value $f_0 = 1/t_0 \approx 6 \times 10^{12}$ Hz from the fitting lines, the slope changes apparently with dc field, indicating the effective barrier $U_0(h_{ac},H_{dc})$ being a function of $H_{dc}$. Plotting $U_0(h_{ac}, H_{dc})$ versus $H_{dc}$ (Fig5) we find a function fitting the eight slopes well is $U_o^{Na}(h_{ac}, H_{dc}) = U_{00}^{Na} H_{dc}^{-m}$, where m=0.86 and $U^{Na}$ represents the energy of YBNCO. Comparing with the data of YBCO (m=0.65), we found that $U_0^{Na} < U_0^0$ for any dc fields, where $U_0^0$ is the effective energy for YBCO. With Na addition, therefore, the flux pinning energy decreased.

Displayed in Fig6 are a summary of the data (symbols) in different amplitudes of ac field $h_{ac}$ in the fixed $H_{dc}$, showing the effect of current density on the effective barriers. Displayed in Fig6 are also $U_o^0(J)$ relationship of YBCO at $H_{dc} = 1$ T for comparison. It is seen that $U^{Na}$ and $U^0$ have similar dependence of J, but the $U^{Na}$ is smaller than $U^0$ at a same J. This is important result showing the negative display of doped Na ion on enhancing flux pinning.

### 3.3. Critical Current Density

Magnetization hysteresis loops were measured and thus magnetic $J_c$ at different temperatures and applied fields were determined based on the expanded Bean model for the Na-doped and Na-free samples, respectively. As an example, shown in Fig.7 is the dependence of $J_c$ on applied field $H_{dc}$ at liquid nitrogen temperature for the samples. From Fig7, two points are clear. First, $J_c$ of YNBCO is lower than of YBCO in the all fields at 77 K. This is directly demonstrated that the doped Na is harmful for flux pinning. Second, a peak in $J_c(H_{dc})$ curve clearly is developed for the Na doped sample, which is also called second peak effect [12]. The second peak has been observed in variety of samples with weak flux pinning. Samples with weak flux pinning are usually a perfect single crystal, a sample with larger anisotropy factor $\gamma$, or a sample at higher temperatures. For examples, the second peaks in magnetization curves have been developed with elevating temperature for $La_{2-x}Sr_xCuO_4$ [13], YBCO [14] and $La_{0.9}Pr_{0.1}Ba_2Cu_{2.62}Al_{0.38}O_y$ [15] single crystals, respectively. Hence, with the temperature increased, the flux pinning is depressed and thus $J_c$ decrease whereas the second peak in magnetization curve gradually was developed with increasing temperature. The second peak in $J_c$ is also an indication that the addition of Na decreases flux pinning of YBNCO.

The above conclusion is further supported by angular dependence of zero resistance temperature $T_{co}$ at the magnetic fields of 1T and 6T respectively [16]. The scaling of the angular dependence of $T_{co}$ with different anisotropic value $\gamma$ in constant fields up to 6T shows that $\gamma = 7$ for YBCO and $\gamma = 11$ for YBNCO, respectively [16]. The results imply that Na substitution give arise to a larger anisotropy in YBNCO crystal. It was suggested that the coupling in Cu-O planes was weakened and the flux pinning was depressed due to the Na addition.

## 4. Summary

The complex ac susceptibility of Na-doped and Na free MTG-YBCO samples was measured systematically as a function of temperature, amplitude and frequency of ac field and dc field to study the flux pinning in the time window of $10^{-4}$s(10 kHz)-$10^{-2}$s(0.1 kHz). The irreversibility lines, effective pinning barriers of flux motion as a function of temperature, dc field and current density were determined for the samples, respectively. By means of magnetic hysteresis loop measurement, the critical current densities of the two samples were also obtained. By comparing these measurements, it is found that with Na addition: 1. The flux pinning energy $U_0$ is decreased.  2. The critical current density $J_c$ is decreased. 3. The irreversibility line is shifted to low temperature. 4. The magnetization curves developed a second peak. 5. The anisotropy became larger. So, it is concluded that the Na addition restrained the superconductivity more than its contribution to flux pinning.

**Acknowledgements**

This work was supported by the Ministry of Science and Technology of China (G1999064602) and National Nature Science foundation of China (NNSFC, 19994016).

Fig captions

Fig. 1. Ac susceptibility- temperature curves at different applied dc fields ($H_{dc}$) of YBNCO. f=2kHz, $h_{ac}$=0.4Gs.

Fig. 2. The irreversibility lines of YBCO and YBNCO. The inset shows the irreversibility lines of YBNCO at different frequency.

Fig. 3. Acs curves at different frequency of YBNCO at $H_{dc}$=0.5T, $h_{ac}$=0.4Gs.

Fig. 4. The plot of $G(T_p)/T_p$ versus $\ln f$ at different dc field of YBNCO at $h_{ac}$=0.4Gs, giving the effective pinning energy U and time scale $t_o$ of flux diffusion.

Fig. 5. The comparison of $U(H_{dc})$ curves between YBNCO and YBCO at $h_{ac}$=0.4Gs..

Fig. 6. The comparison of $U(J)$ curves between YBNCO and YBCO at $H_{dc}$=1T.

Fig. 7. The comparison of $J_c(H_{dc})$ curves between YBNCO and YBCO.

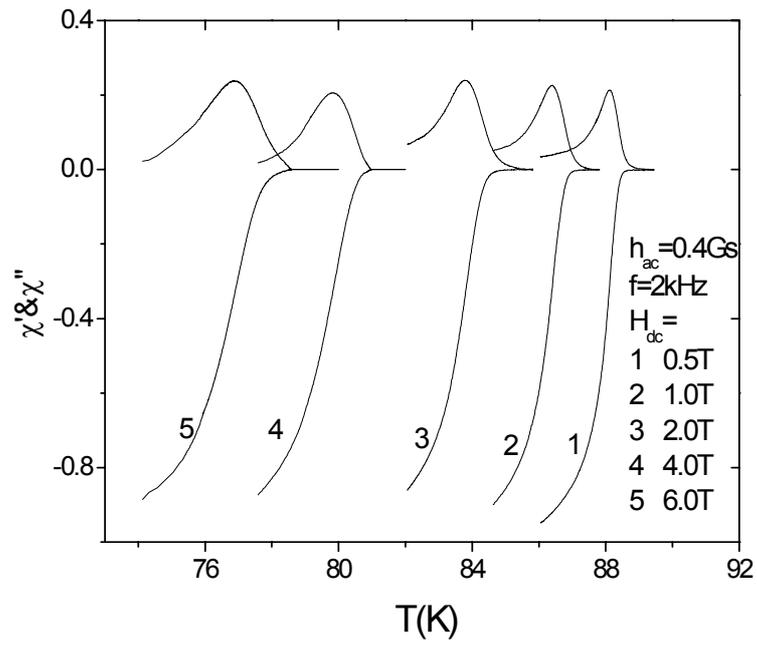

Fig. 1

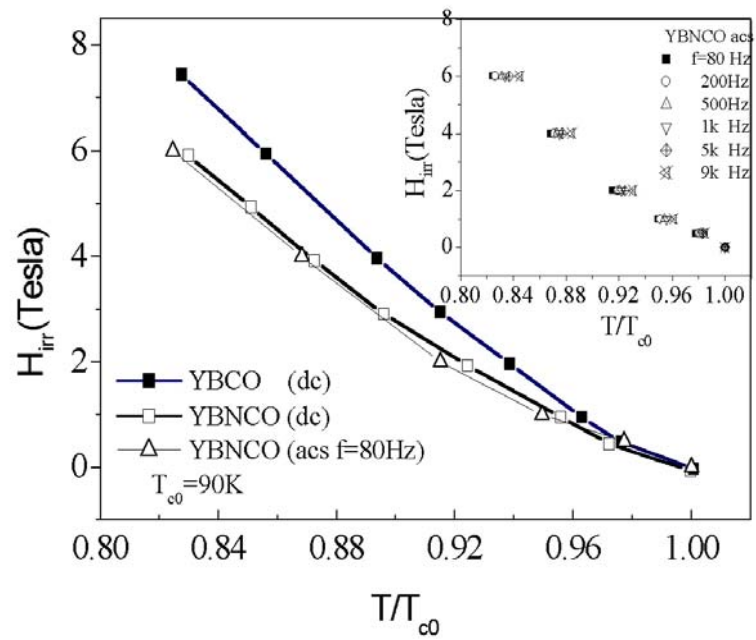

Fig. 2

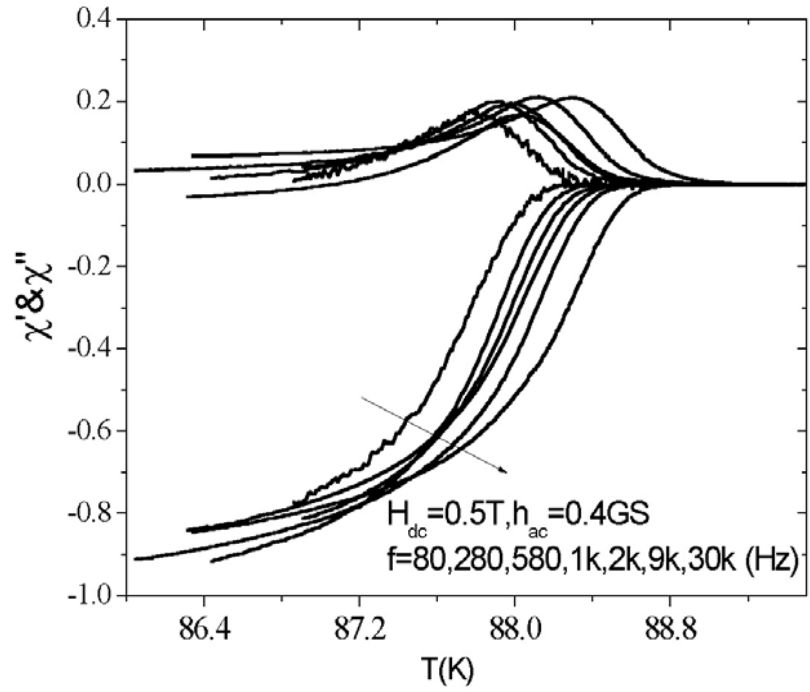

Fig. 3

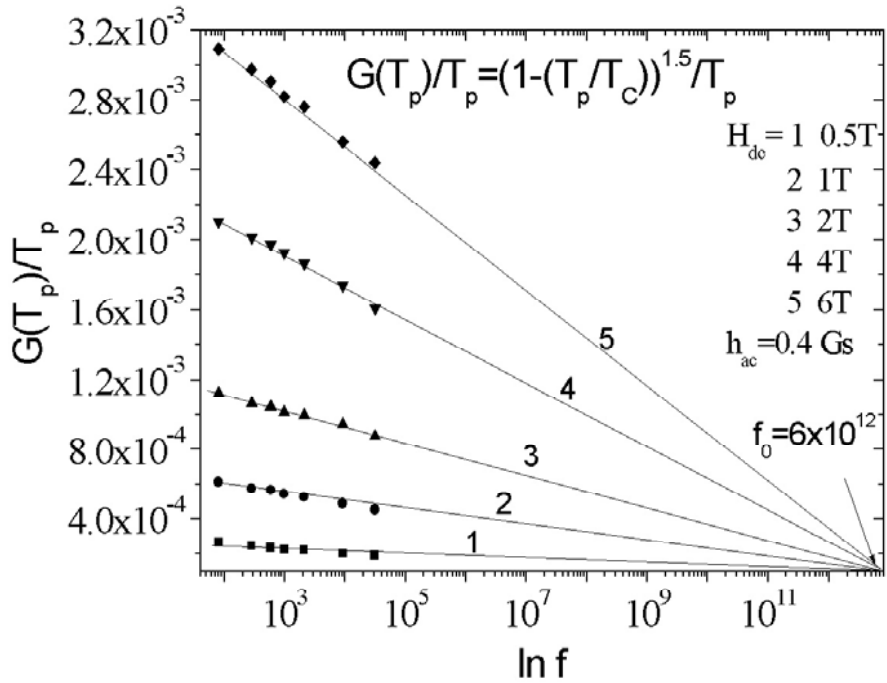

Fig. 4

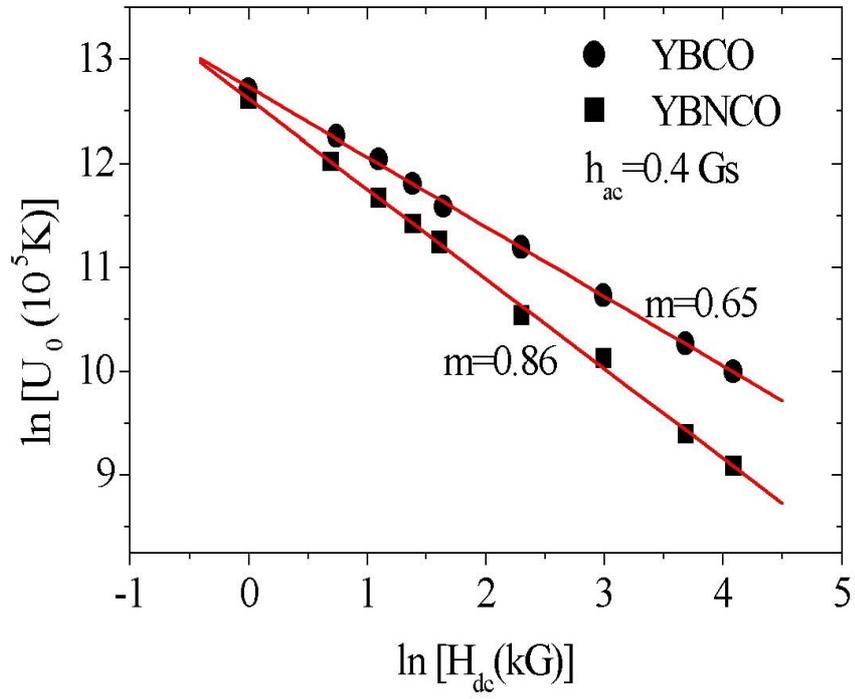

Fig. 5

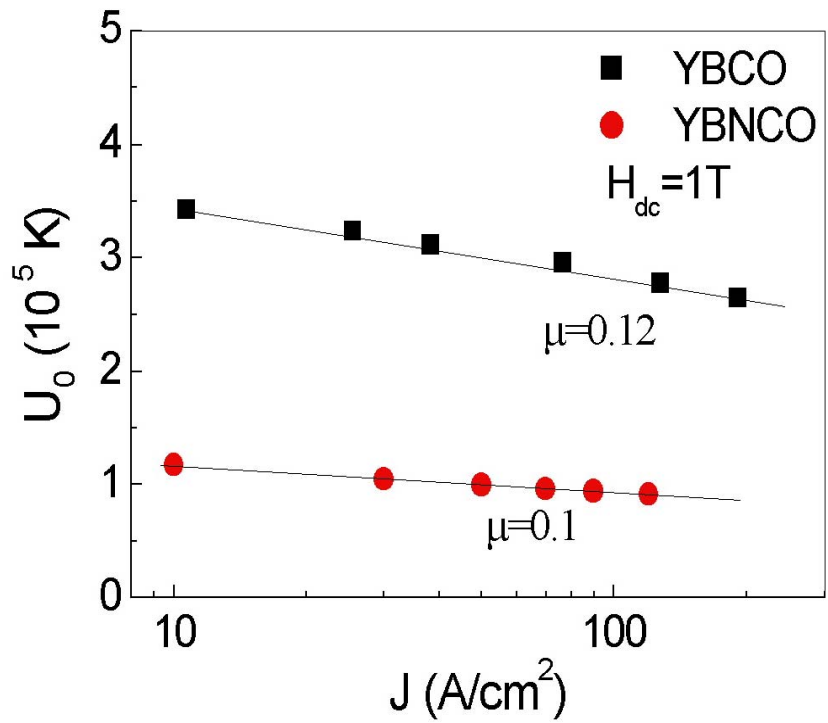

Fig. 6

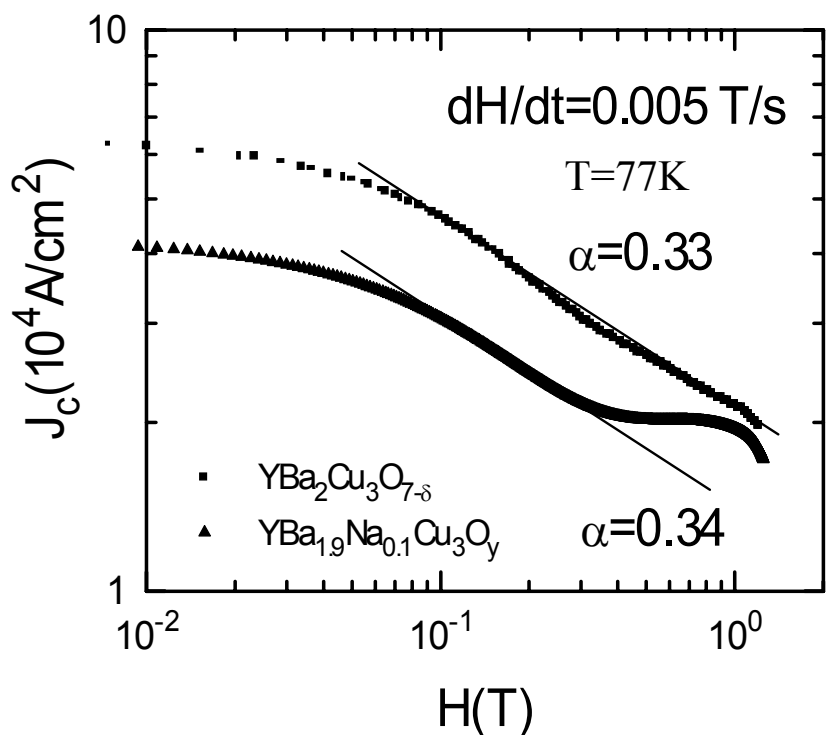

Fig. 7